ASTRONOMY

# Identifying the principal mechanism of star formation feedback

**Chong Li[1]†, Rouyu Li[2,3]†, Keping Qiu[4,5]*, Marko Krco[2], Di Li[6,2]*, Hongchi Wang[1], Xin Lyu[2,3], Yan Duan[2,7], Min Fang[1], Xuepeng Chen[1], Lile Wang[8,9], Haoran Feng[1], Shuinai Zhang[1], Miaomiao Zhang[1], Shaobo Zhang[1], Yue Cao[10], Yao Wang[1], Yan Gong[1], Fujun Du[1], Zhiwei Chen[1], Yang Su[1], Yuehui Ma[1], Yuchen Xing[2,3], Yinghui Zheng[2,3], Tianwei Zhang[11], Shu Niu[1], Chenrui Li[2,3], Jiawei Liu[2,3], Zhijun Zhang[2,3], Yulian Guo[2,3]**

**Feedback from massive stars regulates the formation of next-generation stars and planets and thus shapes the evolution of galaxies, yet observational evidence capable of determining which mechanism dominates the feedback process has long been lacking. Here, we present a large-scale $H_I$ survey toward Orion A. With an unprecedented sensitivity and superior resolution, FAST (Five-hundred-meter Aperture Spherical radio Telescope) reveals a well-defined heart-shaped expanding $H_I$ bubble, enshrouding the well-known "Veil" bubble of ionized gas. The photoionization-driven and wind-driven gas expands at 10.9 and 13.8 kilometers per second, respectively, with a kinetic energy ratio of 1:2 assuming a spherical symmetry and a uniform $H_I$-to-$H_2$ ratio. This first direct comparison within a single bubble shows that radiation does not dominate the feedback process as predicted by classical theories; instead, stellar winds contribute a comparable share of the injected kinetic energy.**

## INTRODUCTION

The feedback process is of central importance in regulating galaxy evolution and star formation. In the absence of feedback, the theoretical star formation rate (SFR) in galaxies is one to three orders of magnitude higher than observed values (*1*, *2*) and there are substantial problems in reproducing the observed stellar initial mass function (IMF) in simulations (*3*, *4*). Meanwhile, feedback-triggered star formation may contribute up to tens of percent to the Galactic stellar population (*5*, *6*). Therefore, feedback plays a key role in regulating star formation in galaxies (*7*). The interstellar medium in the Milky Way and nearby galaxies thus exhibits a widespread distribution of bubbles surrounding massive stars (*8*, *9*). Before star death, the radiation and wind from massive stars dominate feedback processes in galaxies. Regarding their relative importance in momentum injection, however, different theoretical models offer distinct predictions (*10*–*12*), and the direct observational evidence to quantitatively constrain it remains absent to date (*13*).

Approximately 40% of the Galactic neutral hydrogen ($H_I$) mass is shaped by stellar feedback processes (*14*). Their structures and dynamics can thus directly reflect the influence of massive stars on the surrounding environment. During the past half century, $H_I$ shells/bubbles have been widely identified (*15*–*19*). Restricted by data resolution and sensitivity, however, few $H_I$ shells/bubbles are found to be associated with active star-forming regions to allow an assessment of the feedback process. For these reasons, there has been no much progress in our understanding of $H_I$ bubbles over the past decade (*14*). Despite the challenges in directly imaging $H_I$ bubbles enshrouding active $H_{II}$ regions, the gap between the ionization front and dissociation front seen in the most recent JWST images has implied their existence (*20*).

There have been increasing debates regarding the relative importance between the stellar wind and radiation in the feedback process (*21*). The most recent $C_{II}$ observations toward the Orion Nebula, which is part of the FEEDBACK program dedicated to systematically investigating the interaction of massive stars with their surrounding medium (*22*), indicate that the kinetic energy of stellar winds exceeds the expansion energy of the bubble and the radiation pressure is much weaker than the thermal pressure of wind-produced plasma (*23*). The authors thus suggest that the stellar wind dominates the mechanical energy injection to sustain the bubble expansion. However, the x-ray emission, which traces the region influenced by stellar winds, was found to be only concentrated within one half of the bubble. In the other half, the thermal pressure of the photoionized gas is comparable to that of the wind-produced plasma (*24*) and can thus also drive the expansion. Consequently, the finding that the wind energy exceeds the expansion energy alone is insufficient to distinguish which mechanism dominates the bubble's dynamics. To date, there remains no direct evidence capable of unambiguously determining the relative importance between radiation and stellar wind.

## RESULTS

### The most well-defined $H_I$ bubble with FAST

Large-scale $H_I$ spectra with sufficient resolution and sensitivity are crucial for examining feedback effects, as they enable the distinction of various mechanisms underlying the cloud dynamics. The Five-hundred-meter Aperture Spherical radio Telescope (FAST) is recognized as the largest and most powerful single-dish telescope in radio astronomy. The Commensal Radio Astronomy FAST Survey

[1]Purple Mountain Observatory and Key Laboratory of Radio Astronomy, CAS, Nanjing 210033, China. [2]National Astronomical Observatories, CAS, Beijing 100101, China. [3]University of Chinese Academy of Sciences, Beijing 100049, China. [4]School of Astronomy and Space Science, Nanjing University, Nanjing 210023, China. [5]Key Laboratory of Modern Astronomy and Astrophysics (Nanjing University), Ministry of Education, Nanjing, China. [6]New Cornerstone Science Laboratory, Department of Astronomy, Tsinghua University, Beijing 100084, China. [7]Space Engineering University, Beijing 101416, China. [8]The Kavli Institute for Astronomy and Astrophysics, Peking University, Beijing 100871, People's Republic of China. [9]Department of Astronomy, School of Physics, Peking University, Beijing 100871, People's Republic of China. [10]School of Physics and Astronomy, China West Normal University, Nanchong 637002, China. [11]Zhejiang Lab, Hangzhou, Zhejiang 311121, People's Republic of China.
*Corresponding author. Email: kpqiu@nju.edu.cn (K.Q.); dili@mail.tsinghua.edu.cn (D.L.)
†These authors contributed equally to this work.

(CRAFTS) is designed to map the entire sky accessible to FAST with several scientific objectives (*25*). Here, we use the FITS cubes of the CRAFTS $H_I$ Narrow All-Sky (CHINA) Survey, which, as part of CRAFTS, aims to map the local 21-cm emission from ours and other nearby galaxies. The CHINA Survey offers the largest and most sensitive single-dish maps of the $H_I$ Sky in the velocity range from −3000 to +3000 km $s^{-1}$, with a spectral resolution of 0.1 km $s^{-1}$, a spatial resolution of 3 arc min, and a sensitivity of ~25 mK per channel. Upon completion, it will offer an unparalleled view of the $H_I$ content in the local universe.

As such, FAST offers the most sensitive large-scale $H_I$ map of the Orion region to date. We derive the $H_I$ intensity map for the Extended Orion Nebula (EON, also called Veil bubble) in Fig. 1A, which is, to our knowledge, thus far the most well-defined limb-brightening (ringlike) $H_I$ bubble to date. As the nearest and most intensively studied massive star formation region (*26*), the cavity of the expanding bubble has been found to be filled mainly by photoionized gas in the northeastern side and wind-produced plasma in the southwestern one, which produce the radio continuum (*27*) and x-ray emissions (*24*), respectively. The distinct regions influenced by stellar wind and radiation, respectively, along with the excellent FAST image, make the EON one of the very few regions suitable for addressing the issue of which mechanism dominates momentum feedback in active star formation regions.

## Expansion energies driven by photoionization and wind

Figure 1B displays the position-velocity map of the Orion Nebula, which demonstrates a typical expanding structure for the blueshifted (foreground) component. Thanks to the excellent sensitivity of FAST, it is obvious that the maximum expansion velocity of the photoionization-driven region is significantly lower than that of the wind-influenced region (Fig. 1C). We perform the Gaussian decomposition to derive the expansion velocities pixel by pixel for the foreground shell (Materials and Methods) and find that the expansion velocities in the photoionization-driven region are systemically lower than that in the wind-driven zone (Fig. 1D). The $H_2$ column density map derived from SED fitting represents the integrated gas distribution from both the Veil bubble and the Orion molecular cloud, and there is no evidence indicating that the gas mass of the Veil bubble in the northeastern side is overall higher than that in the southwestern side (see the Supplementary Materials). If the mechanical energy injection into the expanding bubble is dominated by photoionization, as suggested by theoretical models (*12*), the northeastern side of the bubble should exhibit a relatively larger expansion velocity.

A previous study speculated that the x-ray–emitting plasma probably fills the whole bubble but is absorbed in the northeastern side, and the bubble expansion is dominated by the stellar wind (*23*). To investigate the distribution of hot plasma, we perform an x-ray absorption analysis toward the phoionization-driven region. The typical hydrogen column density is $\sim 3 \times 10^{21}$ $cm^{-2}$ for the photoionized region (see fig. S3). Assuming that this region is also filled by hot plasma with an intrinsic luminosity equal to that of the wind-driven region, the absorption-corrected x-ray emission should still exceed the background point-source intensity by a factor of 5 to 10 (Materials and Methods). The currently predicted x-ray intensity from the photoionized region represents only a lower limit due to two conservative assumptions: first, that all gas in the column density map lies in the foreground of plasma; second, that the shell thickness of the wind-driven region is negligible ($4 \times 10^{19}$ $cm^{-2}$) compared to that of the photoionized region ($3 \times 10^{21}$ $cm^{-2}$). However, the relatively uniform radius of the expanding bubble indicates no significant difference in shell mass between the photoionized and wind-influenced regions. Therefore, the nondetection of x-ray emission in the photoionized region cannot be attributed to the foreground absorption.

Moreover, a previous study showed that a substantial presence of hot plasma would produce a distinct x-ray–filled cavity within $H_{II}$ regions (*28*). In contrast, the 6-cm radio continuum emission of the EON shows a continuous and uniform distribution of photoionized gas. Therefore, it is unlikely that a large amount of plasma is present within the photoionized region. The $H_I$ kinematics and its coupling with distinct feedback mechanisms demonstrate that the bubble expanding is driven by both radiation and wind, which is inconsistent with either the wind domination suggested by previous observations or the theoretical models of pure photoionization-domination.

In Fig. 2, we display the expansion velocity after the spherical geometric correction, adopting a central position at (R.A., Dec.) = (5 hours 34 min 36 s, −5°33′36″) and a radius of 2.5 pc for the Veil bubble. The median expansion velocities are 10.9 and 13.8 km $s^{-1}$ for clouds influenced by radiation and wind, respectively. Assuming a uniform $H_I$-to-$H_2$ ratio for the foreground ($-10$ to 0 km $s^{-1}$) of the Veil bubble, the median column density of $H_2$ on the southwestern side would be even higher ($3.3 \times 10^{20}$ $cm^{-2}$), reaching 1.3 times that of the northeastern side ($2.6 \times 10^{20}$ $cm^{-2}$). The median expansion energy of the Veil bubble in the wind-driven region is ~2 times that in the photoionization-driven region. Even considering the uncertainties in our energy calculation including the $H_I$-to-$H_2$ ratio (see the Supplementary Materials), geometry, and shell mass, the stellar wind and photoionization play a comparable role in the feedback process. Therefore, neither the stellar wind nor the photoionization can solely dominate the injection of kinetic energy, challenging theoretical models and previous observations that one single mechanism dominants the entire expansion.

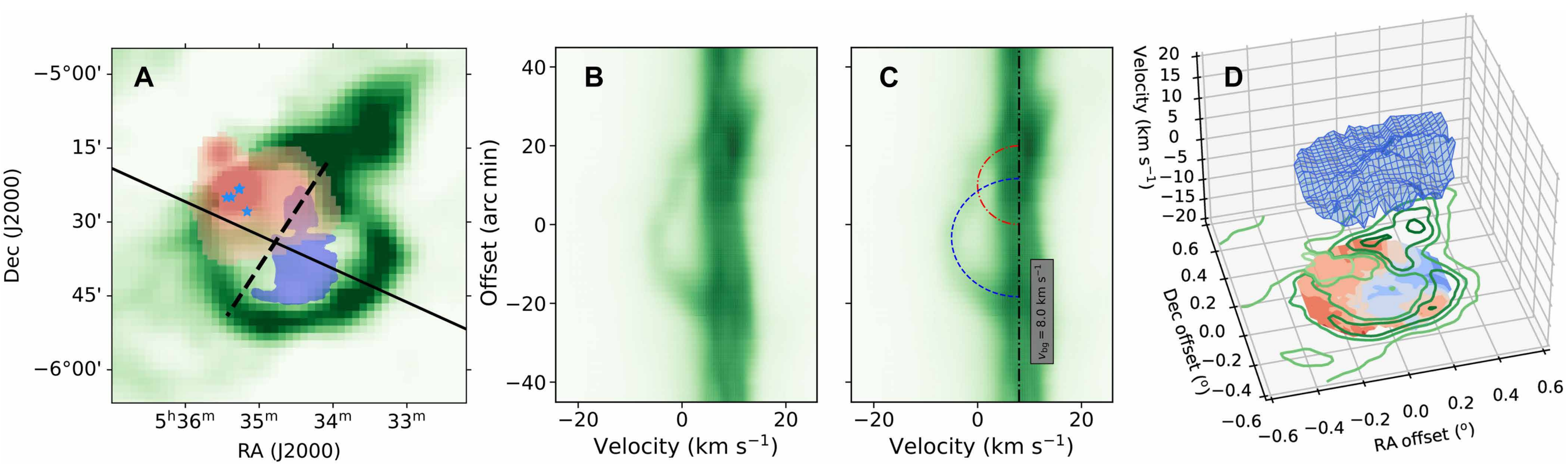

**Fig. 1. Spatial and velocity images of the Veil bubble.** (**A**) $H_I$ integrated intensity map (from 0 to 5 km $s^{-1}$) of the Orion nebula made with the FAST. The pentagrams indicate the Trapezium cluster members. The dashed line divides the bubble into regions dominated by 6-cm radio continuum emission from photoionized gas (shown in orange) and by x-ray emission from wind-produced plasma (shown in blue). h, hours; m, minutes. (**B** and **C**) Position-velocity maps along the solid line in (A) from southwest to northeast, which exhibit separated expanding velocities of the photoionization-driven and wind-driven regions. (**D**) Three-dimensional (3D) (blue grids) and 2D (color) distribution of the radial velocities of expanding components.

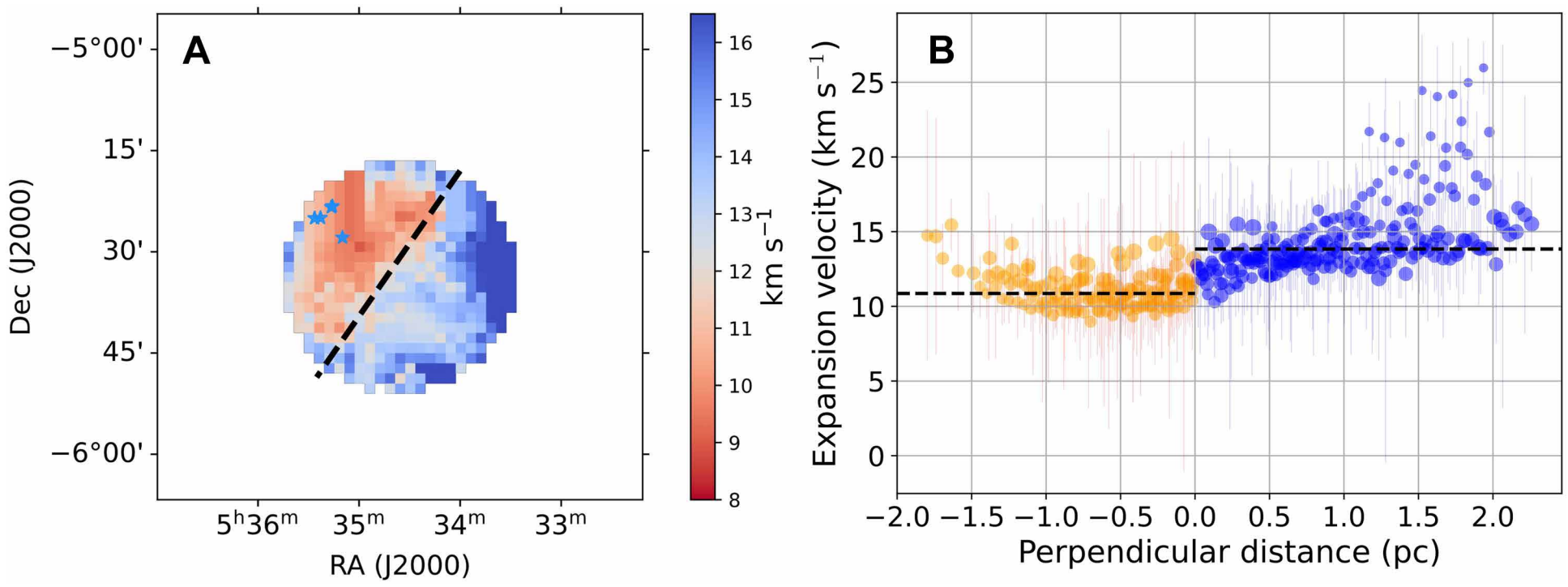

**Fig. 2. Distribution of the expansion velocity of the Veil bubble.** (**A**) Pixel distribution of the corrected expansion velocity for the Veil bubble on the foreground side. Pixels that lack a derived expansion velocity have been interpolated using neighboring values and have been excluded from the right panel. (**B**) Corrected expansion velocity as a function of vertical distance from the dashed boundary in (A). The dashed lines indicate the median velocities for the separated regions. The size of each point is proportional to the peak brightness temperature of the corresponding expanding component.

## Analytic investigations

We analytically evaluate the expanding system driven by dual mechanisms. Constraints on physical parameters can be inferred from the observed quantities on both sides. For the photoionization-driven region, the ionization equilibrium radius of a Strömgren sphere is given by $R_{S0} \approx 9.77 \times 10^{18} Q_{49}^{1/3} n_2^{-2/3} T_4^{0.28}$ cm (*29*). For the Orion Nebula, the ionizing photon luminosity $Q_{49} \equiv Q / \left(10^{49} \text{ photons s}^{-1}\right) \approx 0.56$, the number density of hydrogen nuclei $n_2 \equiv n / \left(10^2 \text{ cm}^{-3}\right) \approx 30$ (*30*), and the temperature $T_4 \equiv T / \left(10^4 \text{ K}\right) \approx 0.8$ (*31*). The initial Strömgren radius of the nebula is calculated to be ~0.3 pc. The propagation timescale of the ionizing front to radius $R_{S0}$ is $\sim 10^2$ years. These values are considerably lower than the size (~2 pc) and dynamic age of the system. The radiative region has completed the initial Strömgren sphere expansion and entered the epoch driven by thermal expansion. The theoretical expansion speed can be estimated from the sound speed inside the ionized gas: $c_s \sim 11$ km $s^{-1}$ (*29*, *32*, *33*), which is approximately equal to the expansion velocity of the photoionization-driven region measured in this work. The dynamic age of the radiative region is $\sim 1.6 \times 10^5$ years.

In the wind-driven region, a large number of bow shocks facing the Trapezium cluster indicate the interaction of stellar winds with the affected regions (*34*). The free-expansion timescale is $t_{\text{free}} \approx R_s / V_w \sim 2 \times 10^3 \text{ years} \times \left(\frac{R}{2\,\text{pc}}\right)\left(\frac{V_w}{10^3 \text{ km s}^{-1}}\right)^{-1}$, which is also considerably shorter than the system age. The system has entered the bubble-filling stage, leaving behind a cavity containing x-ray–emitting plasma with its radius evolving dynamically over time (*29*). We estimate the total gas mass ($M_{H_2} + M_{HI}$) of the Veil bubble to be 1100 solar mass ($M_\odot$) (see the Supplementary Materials). Approximately 1/20 of the mass is contributed by $H_I$ gas, with the initial $H_2$ density of 700 $cm^{-3}$. On the basis of the mechanical luminosity [$L_w \sim 8 \times 10^{35}$ erg $s^{-1}$ (*35*, *36*)] of the wind, we derive an expanding age of $\sim 1.5 \times 10^5$ years (*37*), which matches the dynamic age of the radiative side inferred from its expansion properties. The mass of the hot plasma is 0.07 $M_\odot$, and the wind mass-loss rate is $\dot{M}_{\text{wind}} \sim 10^{-6.7}$ $M_\odot$ $\text{year}^{-1}$, which approximates the lower limit of the stellar wind mass-loss rates for O-type stars. Because of the low density, the radiative cooling of plasma is negligible. Assuming an conservation of kinetic energy, the wind terminal velocity is: $v_{\text{wind}} \sim \left(\frac{M_{\text{shell-wind}}}{M_{\text{plasma}}}\right)^{1/2} \times v_{\text{shell}} \sim 1300 \text{ km s}^{-1}$, which is consistent with the measured velocity of 1650 km $s^{-1}$ (*38*) and a typical value of O-type stars, winds.

On the basis of the above analysis, the radiation and stellar wind feedback from a typical O7V-type star, such as the $\theta^1$ Orionis C, can produce the expanding bubble as observed after a period of evolution. We calculated the proper motion of the Trapezium cluster members to be less than 4′ (*39*), which is smaller than the distance between the cluster and the wind-driven region. Therefore, the shocked gas of stellar wind should have leaked through porous ambient medium, causing the wind-influenced region to be offset from the radiative zone (*40*), thereby providing an opportunity to assess the relative importance of these features in the dynamics. Within the bubble, the thermal pressures of filled ionized gas and plasma are approximately comparable, resulting in a generally balanced and stable internal environment (*24*). A sketch of the expanding $H_I$ bubble driven by wind and radiation is shown in Fig. 3.

# DISCUSSION

## The feedback mechanisms and regulation role of massive stars

Radiation and stellar winds inject momentum into the galactic bubbles through two distinct modes: direct momentum transfer (via radiation pressure and momentum-driven winds) and thermal expansion of heated gas (via photoionization heating and energy-driven winds). For the Veil bubble with a radius of 2 pc, the dynamics of radiation pressure can only make significant contribution when the ionizing photon luminosity reaches $1 \times 10^{51}$ s$^{-1}$ (*41*, *42*), much higher than the photon luminosity of the Trapezium cluster of ~$5 \times 10^{48}$ s$^{-1}$. Conversely, the presence of diffuse x-ray emission provides direct observational evidence that the stellar wind has effectively shock heated the interacting gas to temperatures of ~$10^6$ K, indicating the operation of an energy-driven wind process. Consequently, the two mechanisms with the potential to drive the Veil bubble's expansion are photoionization heating and energy-driven winds.

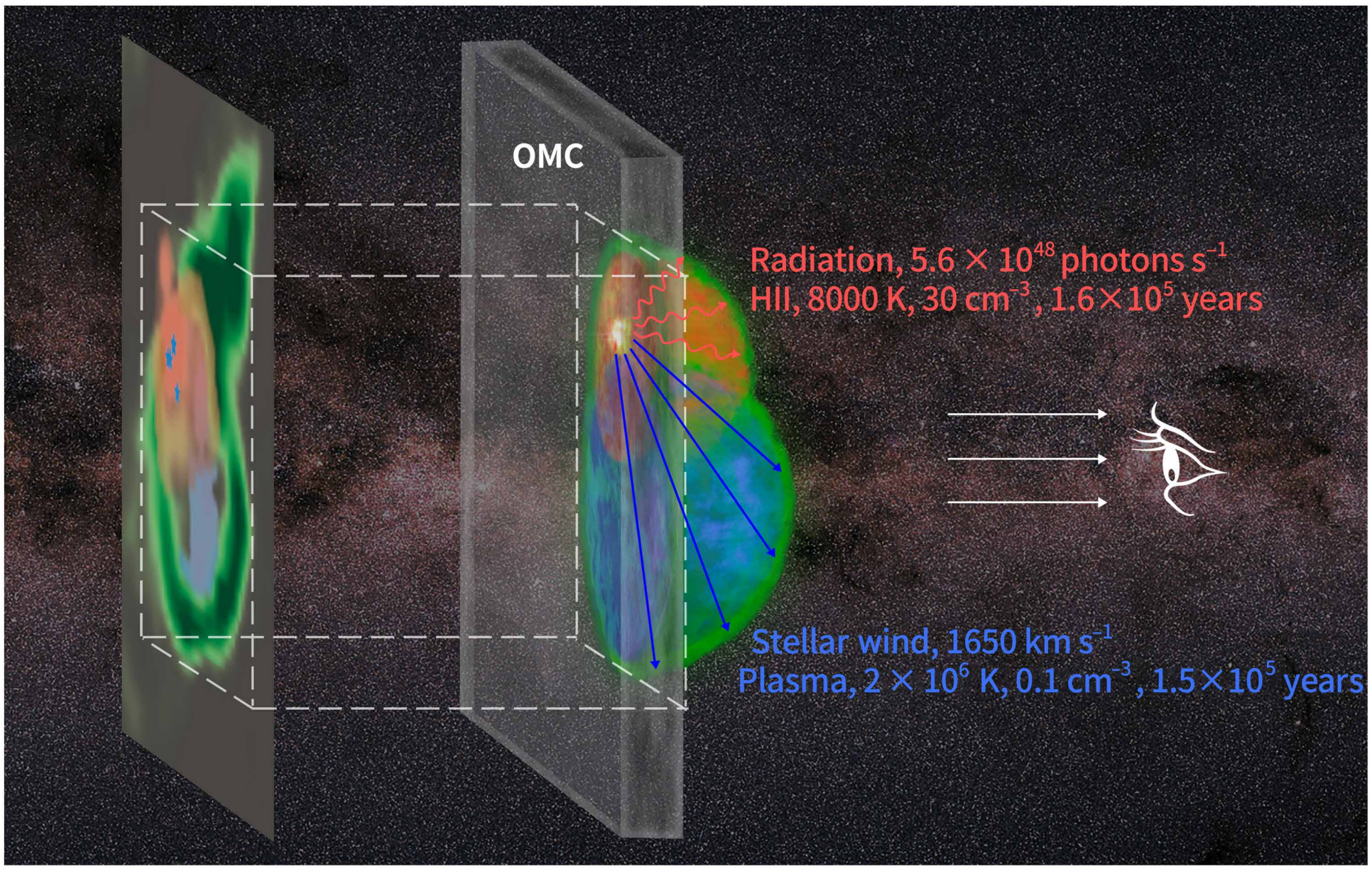

**Fig. 3. Sketch of the expanding bubble.** The scales of radiation-driven and wind-driven regions are based on the corresponding expansion energies, i.e., ~1:2. OMC, Orion molecular cloud.

Traditional models suggest that the kinetic energy of the wind-driven shell would be typically two orders of magnitude higher than the photoionization-driven shell (*43*). Therefore, our observational analysis of the Veil system reveals a picture inconsistent with either pure photoionization-driven or pure wind-driven expansion. Instead, it aligns with the recent and more practical models including turbulence and magnetic field in an aspherical interstellar medium (*11*, *44*, *45*). The distribution of x-ray emission in EON is not well confined within the photoionized gas, which aligns with phenomena in simulations of porous gas. In such environments, plasma can leak through chimneys and may become decoupled from the location of the exciting stars (*40*). During this process, the stellar wind gas loses energy, ultimately leading to comparable feedback effects between the stellar wind and photoionization.

Our observations demonstrate that the expansion system have a dynamic age of ~$1 \times 10^5$ years, indicating the superior role of photoionization and stellar wind in momentum feedback throughout the whole lifetime of massive stars. Although most massive stars end as supernova explosions and release substantial energy, the radiative and wind energy contained are comparable (*9*). Furthermore, supernova theoretically occur 4 to 40 million years after the first star formation in clusters, much delayed than the crossing time for cluster members to convert all gas mass into stars, implying that supernova could not be the dominant feedback mechanism in the local environment (*13*). Similarly in observations, the detectable supernova remnants in our Galaxy is around one order of magnitude lower than that expected from the explosion rate (*46*–*49*), likely because the ISM have already been expelled before the supernova can drive sufficient ambient gas to produce an observable shock (*50*). Although our observations provide valuable constraints on the relative importance of feedback mechanisms during the active star formation phase, systematic multitracer imaging of gas bubbles and further kinematic investigations by FAST $H_I$ surveys are essential to establish its universality spanning various stellar evolution stages.

## MATERIALS AND METHODS

### FAST observations and data reduction

The $H_I$ line emission toward the Orion Nebula was observed under the project of CRAFTS. An L-band Array of 19 feed horns (FLAN) served as the front-end receiver (*25*). The narrow ROACH back-end has a 31.125-MHz bandwidth with 65,536 channels, corresponding to a spectral resolution of 0.476 kHz and a velocity resolution of 0.1 km s$^{-1}$ at 1.420 GHz (*51*). The sky region was scanned along the Right Ascension in drifting scan mode with a scanning rate of 15″/s and a dump time of 0.2 s. A 23.4° rotation between the 19-beam focal plane array and the declination direction was set to facilitate uniform super-Nyquist sampling. The pointing of the telescope reaches a root mean square (RMS) accuracy of 7.9″.

For intensity calibration, a 1.1 K noise from the diode was periodically injected every 200 μs throughout observations. By an absolute measurement of the temperature of the noise dipole, the units of observed data were calibrated into the antenna temperature $T_A$ in kelvin. Subsequently, a Gaussian smoothing kernel was applied to regrid the data, transforming it into a FITS data cube. The beam width is 3′, and the pixel size of the FITS data cube is 1′ × 1′. During observations, the typical system temperature is ~20 K, and the RMS sensitivity is around 25 mK per channel.

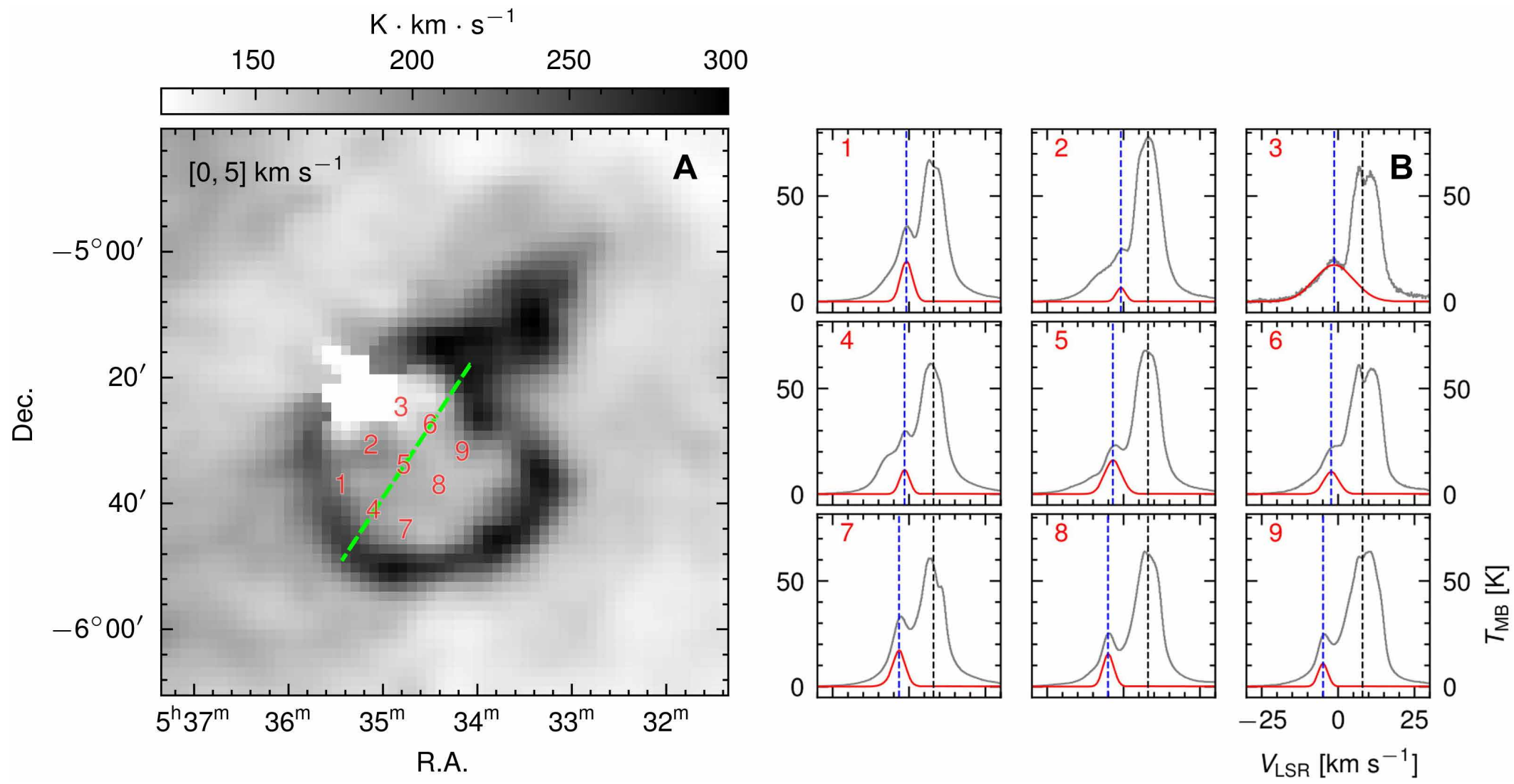

**Fig. 4. Extraction of spectra for expansion component.** (**A**) Integrated intensity map for the velocity interval of [0, 5] km s$^{-1}$, where the red digits from 1 to 9 mark the locations corresponding to the spectra in (B). The green dashed line represents the boundary between the photoionization-driven and the wind-driven regions. (**B**) Spectra taken from the nine pixels shown in (A). The red curves indicate the fitted GCs related to the bubble. The blue dashed lines represent the peak velocities of the expansion components. The black dashed lines represent the background cloud velocity, which is 8.0 km s$^{-1}$.

## Extracting expansion components

To extract the expansion components of the Veil bubble, we applied GaussPy+ (*52*) to decompose the data cube into Gaussian components (GCs) and used the ISMGCC (InterStellar Medium Gaussian Component Clustering) method (*53*) to cluster the blueshifted (foreground) structures. Figure 4 shows the extraction results at nine locations, with points 1 to 3 located in the photoionization-driven region, 4 to 6 located on the boundary, and 7 to 9 located in the wind-driven region. The red curves are recovered from the extracted GCs corresponding to the expansion components. The peak velocities of these recovered spectra match well with the bumps on the raw spectra. By doing so, the central velocity, peak intensity, and linewidth of the expansion component were derived for each pixel.

## Absorption analysis of x-ray emission

The x-ray radiation arises from the plasma heated by wind shock. Previous works have suggested that the stellar wind fills and drives the entire expansion of the Veil bubble. To examine whether the nondetection of x-ray emission in the photoionization region is due to absorption by foreground gas, we perform an x-ray absorption analysis for the EON region based on the model specifically designed for diffuse sources like the X-ray Multi-Mirror Mission (XMM-Newton) (*54*).

In Fig. 5, we derived the x-ray spectrum for the wind-driven region, which is consistent with previous results (*24*). If the photoionized region is also influenced by a stellar wind component, corresponding x-ray emission should be present. Assuming that its intrinsic luminosity is the same as that in the wind-driven region, and adopting a typical hydrogen column density of $N(\mathrm{H}) = 3 \times 10^{21}$ cm$^{-2}$ for the foreground shell (see the Supplementary Materials), the theoretically predicted x-ray intensity after absorption correction would still exceed the background point-source intensity by a factor of 5 to 10. Consequently, there is no convincing evidence that the photoionization-driven region is influenced by wind-produced plasma.

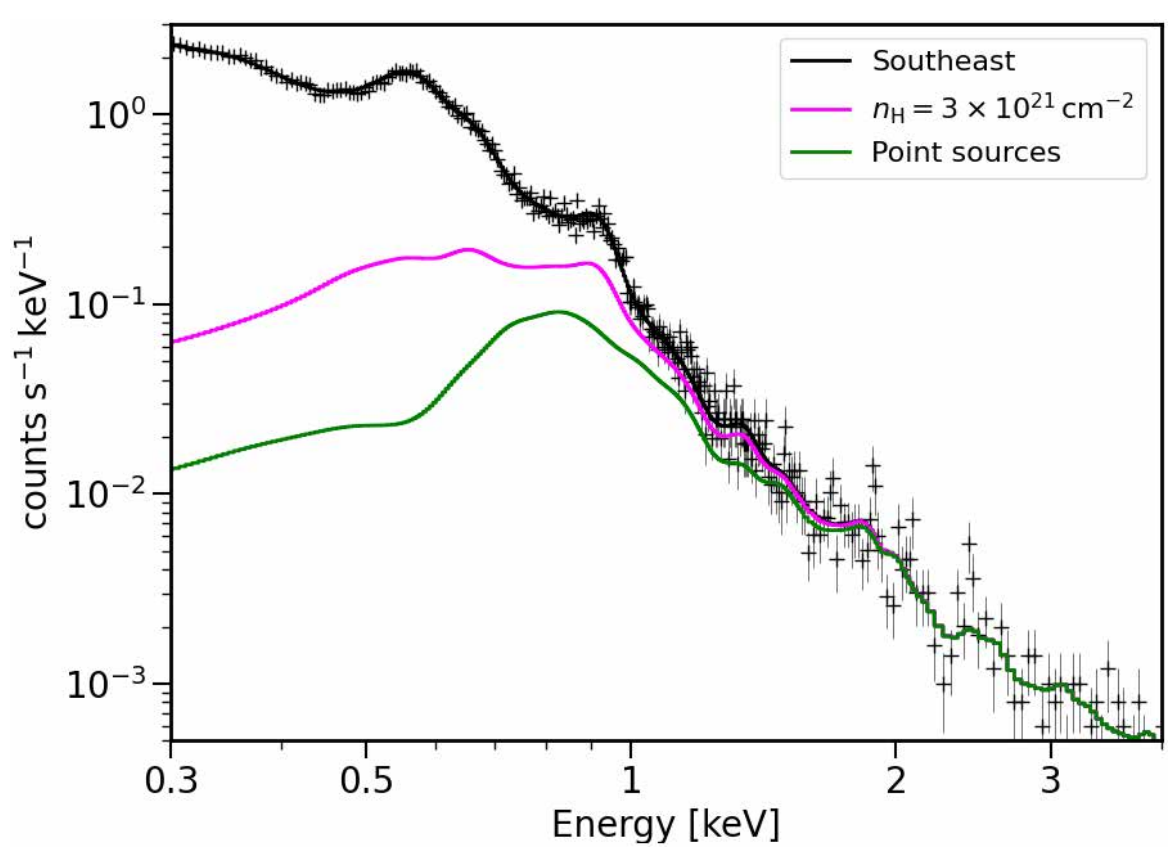

**Fig. 5. X-ray spectra of the EON.** The black and pink spectra correspond to the absorption-corrected southeast and northwest regions, respectively. The error bars associated with each data point represent 1σ uncertainties of counting statistics of the XMM-Newton. The contribution from background point sources is displayed by the green spectrum.

## Supplementary Materials

**This PDF file includes:**
Supplementary Text
Figs. S1 to S6
References

**Acknowledgments:** This work made use of the data from FAST (https://cstr.cn/31116.02.FAST). FAST is a Chinese national mega-science facility, operated by the National Astronomical Observatories, Chinese Academy of Sciences. We thank W. Reich for sharing the data of 6-cm radio continuum emission from Effelsberg. D.L. acknowledges support from the new Cornerstone Science Foundation. This work made use of the x-ray data obtained with XMM-Newton, a European Space Agency (ESA) science mission with instruments and contributions directly funded by ESA member states and the United States (NASA). This work was partly carried out by using $C_{II}$ data from SOFIA. SOFIA is jointly operated by the Universities Space Research Association. **Funding:** This work is supported by the National Natural Science Foundation of China (NSFC) grant nos. 12588202, 12425304, U1731237, 12103025, 12473026, 12303033, 12303032, and 12041305; National Key R&D Program of China grant nos. 2023YFA1608204, 2022YFA1603100, and 2023YFA1608000; and National SKA Program of China grant no. 2025SKA0140100. **Author contributions:** Chong Li and K.Q. designed the project and analysis plan. Chong Li led the scientific analysis, found the distinct expansion properties, and wrote the main text, to which R.L., K.Q., D.L., and H.W. contributed. R.L. analyzed the dynamics of $H_I$ bubbles, drew important scientific conclusions, and wrote part of text, with contributions from X.L.. D.L. was responsible for the CRAFTS survey. M.K. was responsible for the $H_I$ data reduction. H.F. and Y.C. wrote the code to extract parameters of the GCs. M.F., M.Z., Y.W., Y.Gon., Y.M., and S.N. helped to collect ancillary data. L.W. made substantial contributions to the analytical approaches in this work. Shuinai Zhang made the absorption analysis of the x-ray emission. Shaobo Zhang derived the distribution of x-ray emission. Y.D., T.Z., Y.Z., Chenrui Li, J.L., Z.Z. and Y.Guo provided inspiration for the specific discussions involved in the work. All authors participated in the discussion of results and revision of the manuscript. **Competing interests:** The authors declare that they have no competing interests. **Data, code, and materials availability:** All data and code needed to evaluate and reproduce the results in the paper are present in the paper and/or the Supplementary Materials. The $H_I$ data from the CRAFTS $H_I$ Narrow All-Sky (CHINA) Survey can be downloaded from https://doi.org/10.57760/sciencedb.07779. The algorithm (ISMGCC) and parameters used to extract the expansion components can be downloaded from https://zenodo.org/records/19548903. The x-ray raw data from XMM-Newton can be downloaded from https://ui.adsabs.harvard.edu/link_gateway/2008Sci...319..309G/XMM. The data of optical depth $\tau$ at 353 GHz and $\beta$ for Orion are derived by combining the Planck, Herschel, and 2MASS NIR data and can be downloaded from https://cdsarc.cds.unistra.fr/viz-bin/cat/J/A+A/566/A45. The $C_{II}$ data from SOFIA can be downloaded from https://cdsarc.cds.unistra.fr/viz-bin/cat/J/A+A/652/A77. This study did not generate new materials.

Submitted 5 September 2025
Accepted 24 July 2026
Published 4 September 2026
10.1126/sciadv.aec0787

# Identifying the principal mechanism of star formation feedback

Chong Li, Rouyu Li, Keping Qiu, Marko Krco, Di Li, Hongchi Wang, Xin Lyu, Yan Duan, Min Fang, Xuepeng Chen, Lile Wang, Haoran Feng, Shuinai Zhang, Miaomiao Zhang, Shaobo Zhang, Yue Cao, Yao Wang, Yan Gong, Fujun Du, Zhiwei Chen, Yang Su, Yuehui Ma, Yuchen Xing, Yinghui Zheng, Tianwei Zhang, Shu Niu, Chenrui Li, Jiawei Liu, Zhijun Zhang, and Yulian Guo